\documentclass[usenatbib]{mnras}
\usepackage[dvips]{color}
\usepackage{ulem}
\usepackage{longtable}
\usepackage{url}

\usepackage{graphicx}
\usepackage{lscape}

\def\arcmin{\hbox{$^\prime$}}
\def\arcsec{\hbox{$^{\prime\prime}$}}

\def\flux{erg s$^{-1}$ cm$^{-2}$}
\def\lum{erg s$^{-1}$}

\def\aap{A\&A}
\def\mnras{MNRAS}
\def\apjs{ApJS}

\def\apjl{ApJLett}

\def\2s{2S~1553$-$542}

\def\a{$^{\mbox{\small a}}$}
\def\b{$^{\mbox{\small b}}$}

\title[{\it NuSTAR} discovery of a CRSF in 2S~1553$-$542]
{{\it NuSTAR} discovery of a cyclotron absorption line in the transient X-ray pulsar 2S~1553$-$542}
\author[Tsygankov et al.]{Sergey\,S.\,Tsygankov,$^{1}$\thanks{E-mail: stsygankov@gmail.com}
Alexander\,A.\,Lutovinov,$^{2}$
Roman\,A.\,Krivonos,$^{2}$
\newauthor
Sergey\,V.\,Molkov,$^{2}$
Peter J. Jenke,$^{3}$
Mark H. Finger,$^{4}$
and Juri\,Poutanen$^{1,5}$\\
$^1$Tuorla Observatory, Department of Physics and Astronomy, University of Turku,
  V\"ais\"al\"antie 20, FI-21500 Piikki\"o, Finland \\
$^{2}$Space Research Institute of the Russian Academy of Sciences, Profsoyuznaya Str. 84/32, Moscow
  117997, Russia\\
$^{3}$University of Alabama in Huntsville, 301 Sparkman Drive, Huntsville, AL 35899, USA\\
$^{4}$Universities Space Research Association, National Space Science and Technology Center, 320 Sparkman Drive, Huntsville, AL 35805, USA\\
$^{5}$Nordita, KTH Royal Institute of Technology and Stockholm University, Roslagstullsbacken 23, SE-10691 Stockholm, Sweden
}

\begin{document}

\date{Accepted 2015 December 2. Received 2015 November 26; in original form 2015 September 19.}

\pagerange{\pageref{firstpage}--\pageref{lastpage}} \pubyear{2015}

\maketitle

\label{firstpage}

\begin{abstract}
{We report results of a spectral and timing analysis of the poorly
  studied transient X-ray pulsar \2s\ using data collected with the
  {\it NuSTAR} and {\it Chandra} observatories and the {\it Fermi}/GBM
  instrument during an outburst in 2015.  Properties of the source at
  high energies ($>30$ keV) are studied for the first time and the sky
  position had been essentially improved. The source broadband
  spectrum has a quite complicated shape and can be reasonably
  described by a composite model with two continuum components -- a
  black body emission with the temperature about 1 keV at low energies
  and a power law with an exponential cutoff at high
  energies. Additionally an absorption feature at $\sim23.5$ keV is
  discovered both in phase-averaged and phase-resolved spectra and
  interpreted as the cyclotron resonance scattering feature
  corresponding to the magnetic field strength of the neutron star
  $B\sim3\times10^{12}$ G. Based on the {\it Fermi}/GBM data the
  orbital parameters of the system were substantially improved, that
  allowed us to determine the spin period of the neutron star $P =
  9.27880(3)$ s and a local spin-up $\dot P \simeq -7.5\times10^{-10}$
  s s$^{-1}$ due to the mass accretion during the {\it NuSTAR}
  observations.  Assuming accretion from the disk and using standard
  torque models we have estimated the distance to the system
  $d=20\pm4$ kpc. }
\end{abstract}


\section{Introduction}

The transient source \2s\ belongs to the numerous subclass of the
accreting X-ray pulsars with Be optical companions (BeXRP). It was
discovered by {\it SAS-3} observatory during the Galactic plane survey
in 1975 \citep{1976IAUC.2959....2W}.  Later on the transient nature of
the source was established as well as strong pulsations with the
period of 9.3 s and amplitude of $\sim80\%$ were found
\citep{1982IAUC.3667....3K}. Using the same data an orbital period of
the binary system was measured $P_{\rm orb}=30.6\pm2.2$ d and a suggestion
that \2s\ is likely a Be/X-ray binary system has been made
\citep{1983ApJ...274..765K}. It is important to note that no optical
counterpart was directly determined for \2s so far (however see
below).

Second time after the discovery \2s\ came into the view of X-ray
instruments during the outburst in 2007--2008
\citep{2007ATel.1345....1K}.  An intensive monitoring of this outburst
with the {\it RXTE} observatory allowed to improve orbital parameters
of the system and to trace the spectral evolution in the energy range
2.5--30 keV \citep{2012MNRAS.423.3352P}.  Particularly it was shown
that the source spectra in all available intensity states can be well
fitted with the combination of a black-body component (with
temperature varying between 2.5 and 4 keV) and a broken power-law
component. The iron emission line at $\sim6.5$ keV and a strong
photoelectric absorption corresponding to hydrogen column density of
$N_{\rm H}\sim5\times10^{22}$ cm$^{-2}$ were also required by the fit.

The third episode when \2s\ has undergone an outburst was observed in
the beginning of 2015. The increase of the flux seen in the {\it
  MAXI}/GSC data was reported by \cite{2015ATel.7018....1S}. An
estimated starting date of the activity was around 2015 January 28
(MJD 57050).

In this paper we describe results of the comprehensive spectral and
temporal analysis of the high quality data collected by the {\it
  NuSTAR} observatory during the declining phase of this outburst (MJD
57115.48). Main goal was to explore the source properties at high
energies (above 30 keV) for the first time.

\section{Observations and data reduction}

\subsection{{\it NuSTAR} observations}

The {\it Nuclear Spectroscopic Telescope Array (NuSTAR)}
\citep{2013ApJ...770..103H}, launched on 13 June 2012, is the first
orbital X-ray focusing telescope operating at energies above 10
keV. The observatory consists of two co-aligned identical X-ray
telescope systems operating in a wide energy range from 3 to 79 keV with
angular resolution of 18\arcsec (FWHM) and half-power diameter (HPD) of
58\arcsec. Spectral energy resolution of 400 eV (FWHM) at 10 keV is provided
by independent solid state CdZnTe pixel detector units for each
telescope, usually referred as focal plane module A and B (FPMA and
FPMB).

\begin{figure}
\includegraphics[width=0.98\columnwidth,bb=15 200 565 675]{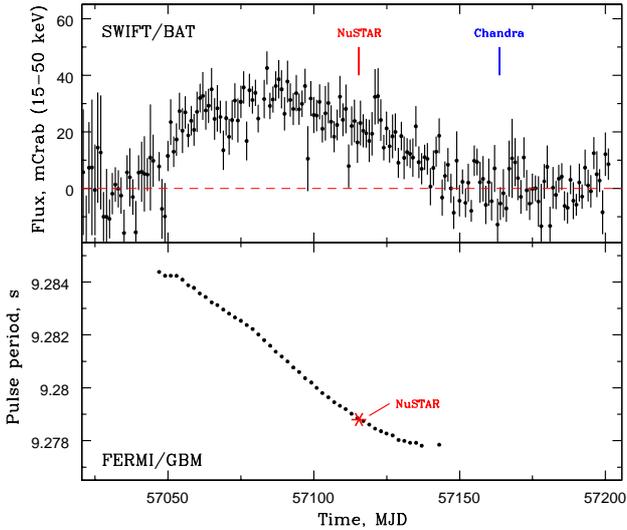}
\caption{{\it Top}: \textit{Swift}/BAT light curve of \2s\ in the 15--50
  keV energy band (black points). Flux is given in units of mCrab (1
  mCrab = $1.4\times10^{-11}$ \flux). Times of the {\it NuSTAR} and
  {\it Chandra} observations are marked. {\it Bottom:} Evolution of the
  pulsar spin period over the outburst as seen by the {\it Fermi}
  Gamma-Ray Burst Monitor (GBM). Asterisk shows corrected for the
  orbital motion period measured by {\it NuSTAR} (see text for the
  details).}\label{fig:lc}
\end{figure}

\begin{figure}
\includegraphics[width=0.998\columnwidth,bb=45 275 545 515,clip]{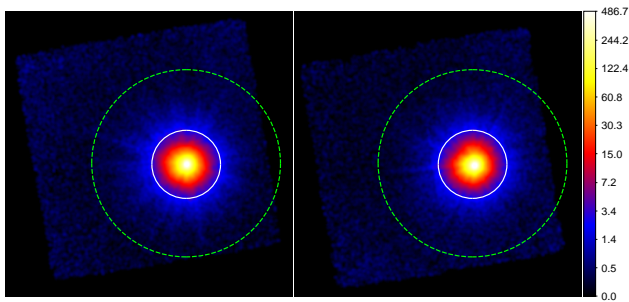}
\caption{Exposure-corrected sky images of \2s\ from FPMA (left) and
  FPMB (right) in $3-79$ keV energy band. The images have been
  smoothed by a Gaussian kernel with 9\arcsec\ width and color-coded in
  logarithmic scale for convenience. The color bar on the right
  indicates the units of the images expressed in $10^{-4}$ cts
  pix$^{-1}$ s$^{-1}$. The solid white circle (120\arcsec\ radius)
  denotes region for the source extraction. The {\sc nuskybgd}
  background model has been calibrated in the area outside the green
  dashed circle (330\arcsec\ radius).}\label{fig:ima}
\end{figure}

{\it NuSTAR} performed a TOO observation of \2s\ in the declining
phase of the outburst (MJD 57115.4834, see Fig.~\ref{fig:lc}) with a
total exposure time of 27 ks (Obs. ID 90101002002). The source was
relatively bright demonstrating net count rate of about 20 cts
s$^{-1}$ on both FPMA and FPMB. We did not notice any special issues
related to the high count rate of the source in the following
analysis.
The source covered large area of the FOV (Fig.~\ref{fig:ima}) because
the {\it NuSTAR} PSF has wide wings
\citep{2013ApJ...770..103H,2014SPIE.9144E..1QA}.  The data were
reduced using the NuSTARDAS pipeline version v1.4.1 (28 May 2014) and
CALDB version 20150612.

In order to perform a barycentric corrections it is necessary to know
the source position with a good accuracy.  Observations with the {\it
  Chandra} observatory, performed on 2015 May 21 (MJD 57163.68;
ObsID. 17662), allowed us to measure it as R.A.= 15$^{\rm h}$57$^{\rm
  m}$48\fs3, Dec.= $-$54$^\circ$24\arcmin53\farcs1 (J2000)
with an 1\arcsec\ uncertainty (90\%). Note, that these {\it Chandra}
observations as well as observations with the {\it Swift}/XRT
telescope were used to determine the optical counterpart in the system
and to measure the absorption value \citep{lut_chasalt}. The latter is
important for the analysis of the {\it NuSTAR} data as it works above
3 keV and not very sensitive to the small or moderate absorption
values. We would like to remind here that a quite high photoabsorption
($4-5\times10^{22}$ cm$^{-2}$) was required by
\citet{2012MNRAS.423.3352P} to describe the source spectra, obtained
with the {\it RXTE} observatory (it operates also above 3 keV). Such
high values are quite atypical for BeXRPs, whose spectra do not
demonstrate usually a strong absorption. Our analysis of {\it Chandra}
and {\it Swift}/XRT data in soft X-ray band shows lower
photoabsorption value in the source spectrum around
$\simeq2.3\times10^{22}$ cm$^{-2}$ \citep{lut_chasalt},
which was fixed in the following spectral analysis of the {\it NuSTAR}
data. Note, that it is only slightly higher, than the estimations of
the galactic interstellar absorption in this direction
$\sim1.6-1.9\times10^{22}$ cm$^{-2}$ \citep{1990ARA&A..28..215D,LAB}.

We extracted spectra using {\sc nuproducts} script provided by the
NuSTARDAS pipeline. The source spectrum was extracted within
120\arcsec\ aperture around the source position as shown in
Fig.~\ref{fig:ima}, which constitutes ~92\% of PSF enclosed energy
\citep[see, e.g., ][]{2014SPIE.9144E..1QA}. The {\it NuSTAR}
background varies across the FOV due to the different stray light (also
called ``aperture'') background components: individual bright X-ray
sources, isotropic extragalactic Cosmic X-ray Background (CXB) and
Galactic X-ray Ridge Emission \citep[GRXE; ][]{2014ApJ...792...48W}. The field
around \2s\ does not contain stray light from nearby Galactic sources,
but it has CXB and GRXE components in the background,
because \2s\ is located in the Galactic plane ($l^{II}$=$-$32.056,
$b^{II}$=$-$0.857). We used suite of IDL routines {\sc nuskybgd}
\citep{2014ApJ...792...48W} to model all the known background
components (instrumental, CXB and GRXE) in the source-free region
outside the green dashed circle (R=330\arcsec) shown in Fig.~\ref{fig:ima}. This
model was utilized to estimate background spectrum at the position of
the source.

\subsection{{\it Fermi}/GBM and {\it Swift}/BAT observations}

\textit{Fermi} Gamma Ray Burst Monitor (GBM) is an all sky monitor
whose primary objective is to extend the energy range over which
gamma-ray bursts are observed in the Large Area Telescope (LAT) on
\textit{Fermi} \citep{2009ApJ...702..791M}.  GBM consists of 12 NaI
detectors with a diameter of 12.7 cm and a thickness of 1.27 cm and
two BGO detectors with a diameter and thickness of 12.7 cm.  The NaI
detectors have an energy range from 8 keV to 1 MeV while the BGO's
extend the energy range to 40 MeV.

The \textit{Swift} Burst Alert Telescope (BAT) is a hard X-ray monitor
that has a field-of-view of 1.4 steradians and its array of CdZnTe
detectors are sensitive in 15--150 keV range
\cite{2013ApJS..209...14K}.  It is a coded aperture instrument with a
detector area of 5200 cm$^{2}$.  We use the BAT transient monitor
results\footnote{\url{http://heasarc.gsfc.nasa.gov/docs/swift/results/transients/}}
(15--50 keV), provided by the BAT team in order to model the torque
imparted to the neutron star by the accreted material.

GBM Channel 1 (12--25 keV) CTIME data
from 57020--57145 MJD is binned to 250 ms and fit to a semi-empirical
background model.  The background model is subtracted and pulsed flux
and frequencies for 2S 1553-542 are extracted from the data by
modeling its Fourier components \citep{2012ApJ...759..124J, 1999ApJ...517..449F}.

\section{Results}

As can be seen from Fig. \ref{fig:lc}(a) total duration of the analysed
outburst is around 3.5 months covering more than 3 complete binary
orbital cycles. Such a duration is typical for Type II outbursts from
BeXRPs. These events are caused by the non-stationary increase of the amount of
matter in the Be circumstellar disc. Peak
luminosities can be much higher than $10^{37}$ \lum~ \citep[for a review,
  see][]{2011Ap&SS.332....1R}.

\subsection{Orbital parameters}
\label{sec:orb}

We have determined an orbital model for 2S 1553$-$542 using the
\textit{Fermi}/GBM and the \textit{Swift}/BAT data.  Typically, after
correcting the pulse arrival times for Earth's motion, one can use the
doppler boosted frequencies of the pulses to fit a model for the
binary system.  Material accreted onto a neutron star surface or
collected in an accretion disk threaded by the neutron star's magnetic
field transfers angular momentum to the neutron star.  Disentangling
this intrinsic spin-up from the orbital signature is challenging.  A
solution to this problem is to model the intrinsic spin-up using a
proxy for the system's X-ray luminosity.  The luminosity is a function
of mass accretion which is related to the torque imposed on the
neutron star.  \citet{1977Natur.266..123R} showed that, at high
luminosity, the intrinsic spin-up, $\dot{\nu}$, is proportional to
$L^{6/7}$ when accretion is mediated through a disk.  The
proportionality constant is a function of mass, radius, moment of
inertia, distance and magnetic field of the neutron star along with a
few parameters describing accretion and emission efficiency.  When
connecting multiple outburst with the same torque model, it is
necessary to include a spin-down term to account for angular momentum
losses during quiescence.

A search for frequency and frequency rate is performed using pulse
profiles from GBM folded over a two day interval.  Each frequency
epoch is chosen as the mean exposure-weighted observation time.  The
epochs are barycentered using the JPL Planetary ephemeris DE200
\citep{1990A&A...233..252S}.  BAT survey data (15--50 keV) for 2S
1553$-$542 are used as a proxy for the source luminosity and to model
the intrinsic spin-up rate.  In order to eliminate under-constrained
and over-constrained rates, only BAT rates with errors greater than
$10^{-3}$ and less than 0.05 are used.  This spin-up model along with
the line of sight delay associated with the binary orbit from
\citet{1981ApJ...247.1003D} is used to model the barycentered arrival
times.  Minimization of the $\chi^{2}$ fit of the barycentric
frequencies and the BAT rates is performed using the
Levenberg-Marquart method and is of the form:
\begin{eqnarray}
\chi^{2} =& \sum\nolimits_{i}(f_i-(\dot{\nu_i}(1-\beta_i))^2  &; i=0, n-1\\
              & + \sum\nolimits_{i}(X_i - \dot{\nu_i} / m)^2         &; i = 0,n-2
 \end{eqnarray}
where $\beta_i$ is the orbital redshift factor at time $t_i$ which is a function of the orbital elements, $f_{i}$ is the measured barycentric frequency at time $t_{i}$ which is the frequency epoch of the search interval $i$.  Each epoch is chosen as the mean exposure-weighted observation time. 
 $X_i$ is the average value of $R^{6/7}$ between $t_i$ and $t_{i+1}$ and $R$ is the BAT rate.  
 The model parameter $\nu_i$ is the orbitally corrected frequency at time $t_i$ and $\dot{\nu_i}$ is $(\nu_{i+1} - \nu_i)/(t_{i+1} - t_i)$. 

The updated orbital model is used in a new search for frequencies and
frequency rates recursively until the orbital solution becomes
stationary.  The final fit resulted in a $\chi^{2} = 66.3$ with 38
d.o.f.  Variability in the pulse profile within the two day integration
interval contributed to errors in the measured frequency.  In
addition, changes in the emission beam within the integration interval
or systematically throughout the outburst is expected to produce
systematic errors in the BAT rates used to model the spin-up.  In
order to adjust the errors on the orbital parameters to account for
these issues, the errors are increased by 1.35 which results in a
orbital fit with a reduced $\chi^{2} \sim 1$.

\begin{table}
\begin{center}
\caption{Orbital Ephemeris for 2S 1553$-$542.}
\begin{tabular}{lcc}
\hline\hline

Orbital Period        &  31.303(27) & days   \\  
T$_{\pi/2}$                     &   2457089.421(19) & days    \\
$a_x \sin i$                &   201.25(84)& lt-sec     \\
Eccentricity             &    0.0351(22) &     \\
Longitude of periastron  &   163.4(35) & degrees      \\
$\chi^{2}$                      &    66.3 / 38 d.o.f. &              \\
\hline
\end{tabular}
\end{center}
\end{table}

This ephemeris is orbitally phase connected, within $2.7\pm2.3 $ days,
to the orbit determined in \citet{2012MNRAS.423.3352P} from the 2008 outburst.
Spin down between the 2008 outburst and 2015 outburst is $\sim 6.5$
cycles per day.

The frequency history, pulsed flux and ephemeris for this source and
all other sources monitored by the GBM Pulsar Monitoring team may be
found at the GBM pulsar
website\footnote{\url{http://gammaray.msfc.nasa.gov/gbm/science/pulsars.html}}.

\subsection{Pulse period and profile}

To determine the mean pulse period and its uncertainty using the {\it
  NuSTAR} observation we took the barycentered light curve in the wide
energy band (3--79 keV) and based on that generated a set of $10^4$ trial
light curves. For each light curve we found a pulse period
and got a distribution of these trial periods. Individual pulse
periods in each light curve were obtained using {\sc efsearch}
procedure from FTOOLS package. The mean value of this distribution was
taken as proper pulse period of \2s\ and the standard deviation -- as
the corresponding $1\sigma$ uncertainty \citep[see,
  e.g.,][]{2013AstL...39..375B}.

Following this procedure we obtained the spin period $P_{\rm
  spin}=9.282204(2)$ s in the raw data. This corresponds to the
intrinsic spin period value $P_{\rm spin}=9.27880(3)$ s after the
correction for the orbital motion using orbital parameters listed in
Section \ref{sec:orb}. An uncertainty for the intrinsic period value
was calculated using Monte-Carlo simulations taking into account
uncertainties in the orbital parameters and using standard relations
between the pulse period and ephemeris. Spin period from the {\it
  NuSTAR} data is shown in the lower panel of Fig. \ref{fig:lc} by the
asterisk. Interestingly, it is roughly equal to the period value at
the moment of the source discovery around 40 years ago. In spite of
significant spin-up observed during the outburst episode (see bottom
panel of Fig.\ref{fig:lc}) the period constancy over decades is not
surprising due to rarity of such accretion events and slow spin-down
between them. Very similar picture is seen in many other rarely
bursting transient BeXRPs, like 4U~0115+63, V~0332+53, A~0535+262 and
others.$^{\textcolor{blue}2}$

Pulse profile shape carries useful information about geometrical
properties of the emitting area at the neutron star surface, whereas
its dependence on energy reflects physical properties of
matter--radiation interaction. Following \cite{2015ApJ...809..140K}
instead of analysing light curves from modules FPMA and FPMB
independently we combined them in order to get better statistics.

\begin{figure}
\includegraphics[width=0.98\columnwidth,bb=70 140 500 710]{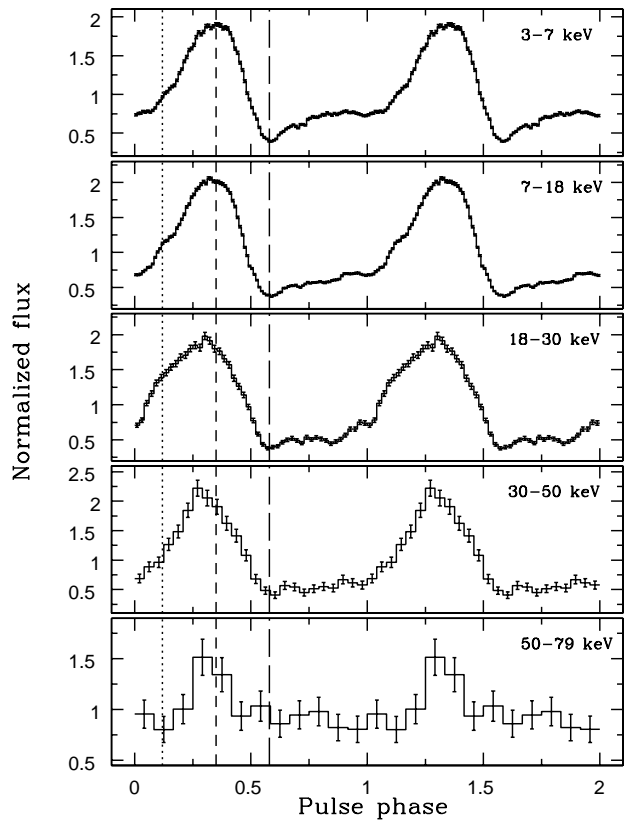}
\caption{Pulse profiles of \2s\ as seen by {\it NuSTAR} in five
  different energy bands 3--7, 7--18, 18--30, 30--50, and 50--79 keV
  normalized by the mean flux in each band (from top to bottom). The
  profiles are plotted twice for clarity. Vertical lines show the
  positions of the most prominent features: main maximum (short-dashed
  line), main minimum (long-dashed line) and energy-dependent wing
  (dotted line).}\label{fig:pprof}
\end{figure}

Pulse profile has a single-peak shape with a barely noticeable
dependence on energy. Figure \ref{fig:pprof} illustrates evolution of
the normalized for the mean flux pulse profile over five energy bands:
3--7, 7--18, 18--30, 30--50, and 50--79 keV (from top to bottom).
Three main features can be distinguished: main maximum, main minimum
and energy-dependent wing, appearing in 18--30 keV energy band. In the
figure these features are shown by short-dashed line, long-dashed line
and dotted line, correspondingly.

\begin{figure}
\includegraphics[width=0.98\columnwidth,bb=70 140 500 695]{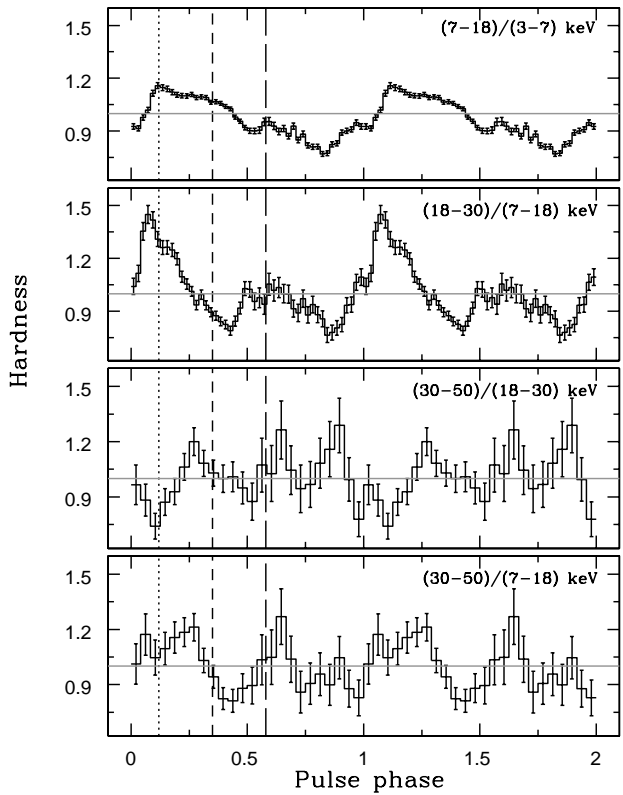}
\caption{Hardness ratios of \2s\ pulse profiles in different energy
  bands (specified in each panel). Vertical lines show the positions
  of main features in the pulse profile (see
  Fig. \ref{fig:pprof}). Grey horizontal line represents the hardness
  ratio of unity. }\label{fig:hard}
\end{figure}

Most obviously pulse profile dependence on energy is illustrated by
hardness ratios shown in Fig. \ref{fig:hard}. Quite complicated
structure of all four ratios is mainly defined by energy dependence of
the main peak. Namely, an additional component is appearing in 18--30
keV energy band at phases $\sim$0.1--0.2 (shown by dotted
line). Interestingly, this component disappears at higher
energies. This is clearly seen from the decrease of the hardness
ration below the level of unity in 30--50/18--30 keV in contrast to a
significant excess in 18--30/7--18 keV and almost ratio of unity in
30--50/7--18 keV bands.

\begin{figure}
\includegraphics[width=0.98\columnwidth,bb=45 310 515 720]{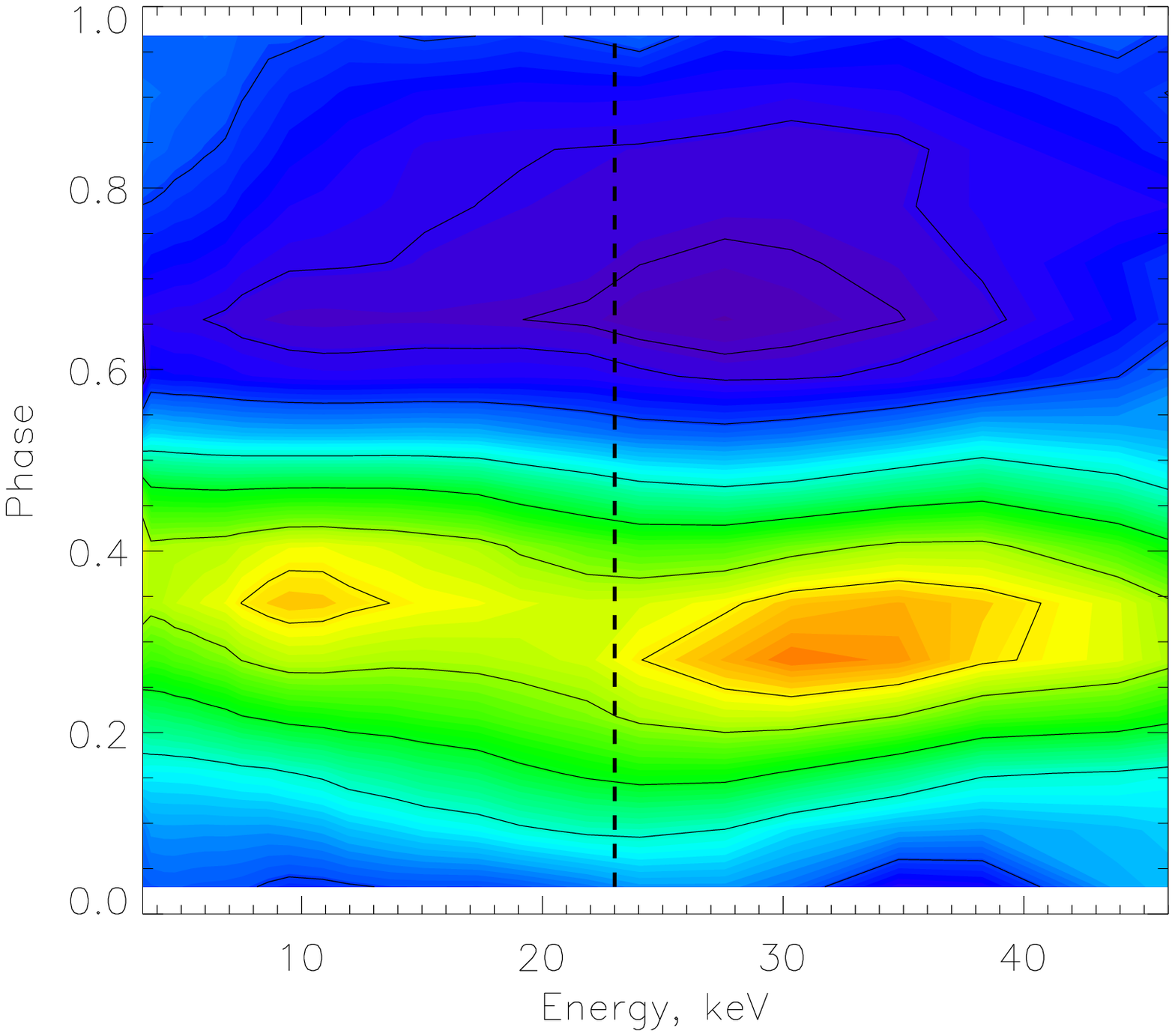}
\caption{Normalized flux map in the ``energy -- pulse phase''
  coordinates. Dashed vertical line shows the position of the
  cyclotron line centroid energy. Solid contours represent
  levels of equal normalized flux at 0.45, 0.5, 0.7, 1.1, 1.4, 1.7,
  1.9.}\label{fig:2dpprof}
\end{figure}

\begin{figure}
\includegraphics[width=0.98\columnwidth,bb=20 270 515 675]{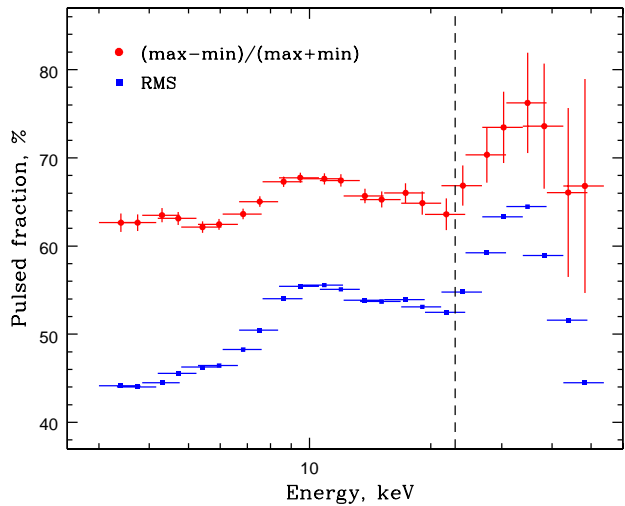}
\caption{Dependence of the pulsed fraction of \2s\ on energy obtained
  using two different approaches (see text for the details) based on
  the {\it NuSTAR} data. Dashed line shows the position of CRSF
  discovered in the source spectrum.}\label{fig:ppfr}
\end{figure}

This behaviour is resulted in a wave-like structure of the pulse
profile shape as function of energy and is clearly seen in
Fig. \ref{fig:2dpprof}. Dashed line shows the position of the
discovered cyclotron line (see Section~\ref{sec:spec}). To construct this
figure we used pulse profiles normalized by the mean flux value in
each energy band, which were chosen to be the same as in
Fig. \ref{fig:ppfr}.

The pulsed fraction has virtually constant value around 60--70 per
cent between 3 and 20 keV, whereas at higher energies it shows a
non-monotonic dependence on energy (see Fig. \ref{fig:ppfr}). The
observed behaviour of the pulsed fraction can be interpreted as local
minimum around $\sim25$ keV superimposed on the gradual increase with
energy, typical for the majority of X-ray pulsars
\citep{2009AstL...35..433L}. Note, that only half of all energy bands
here are statistically independent, however such representation (equal
to a running average) reveals features in the most evident way.  At
energies above $\sim40$ keV one can observe another decrease of the
pulsed fraction. Its possible physical explanation is an influence of
the second harmonics of the cyclotron line, which is expected at these
energies. However, as can be seen from Fig. \ref{fig:avspec} the
signal from \2s\ becomes background dominated above 40--50 keV and
precise determination of the pulsed fraction depends strongly on the
background model. Therefore, the second decrease in the pulsed
fraction should be interpreted with caution.

We used two different definitions of the pulsed fraction to avoid any
possibly biases due to pulse profile shape or statistics.  Standard
definition of the pulsed fraction is
$\mathrm{PF}=(F_\mathrm{max}-F_\mathrm{min})/(F_\mathrm{max}+F_\mathrm{min})$,
where $F_\mathrm{max}$ and $F_\mathrm{min}$ are maximum and minimum
fluxes in the pulse profile, respectively. Defined this way
pulsed fraction is shown by red circles in
Fig. \ref{fig:ppfr}. Another approach to characterize flux
variations over the pulse is to use the relative Root Mean Square
(RMS), which can be calculated using the following equation:

\begin{equation}\label{rms}
RMS=\frac{\Big(\frac{1}{N}\sum_{i=1}^N(P_i-\langle P\rangle)^2\Big)^{\frac{1}{2}}}{\langle P\rangle},
\end{equation}
where $P_i$ is the background-corrected count rate in a given bin of
the pulse profile, $\langle P\rangle$ is the count rate averaged over the pulse
period, and $N$ is the total number of phase bins in the profile.  This
definition gives significantly lower absolute value of the pulsed
fraction, however all features described above have the same form and
are located at the same energies.

\subsection{Spectral analysis}
\label{sec:spec}

The main purpose of the performed {\it NuSTAR} observations was
searching for the cyclotron resonant scattering feature (CRSF) in the
source spectrum, measuring its parameters in the case of the discovery
or putting strict limits on its presence. The {\it NuSTAR} observatory
is perfectly suitable for these purposes, as it covers uniformly the
energy range from 3 to 79 keV (without a necessity to combine
different instruments with different response matrixes, like it was
for the {\it RXTE} or {\it INTEGRAL} observatories) and has an
unprecedented sensitivity in the energy range 10--50 keV, where most
of known cyclotron absorption lines were detected \citep[see,
  e.g.,][for a recent review of high mass X-ray
  binaries]{2015arXiv150503651W}.  Moreover, good timing capabilities
of {\it NuSTAR} allow to perform the same analysis also for pulse
phase resolved spectra, as sometimes cyclotron features are detected
only for limited pulse phase intervals and not seen in the pulse
averaged spectra. In the following spectral analysis of the source
emission we used fixed hydrogen column density $2.3\times10^{22}$
cm$^{-2}$ \citep{lut_chasalt}.

\subsubsection{Pulse-phase averaged spectroscopy}

\begin{figure}
\includegraphics[width=0.98\columnwidth,bb=40 280 550 695]{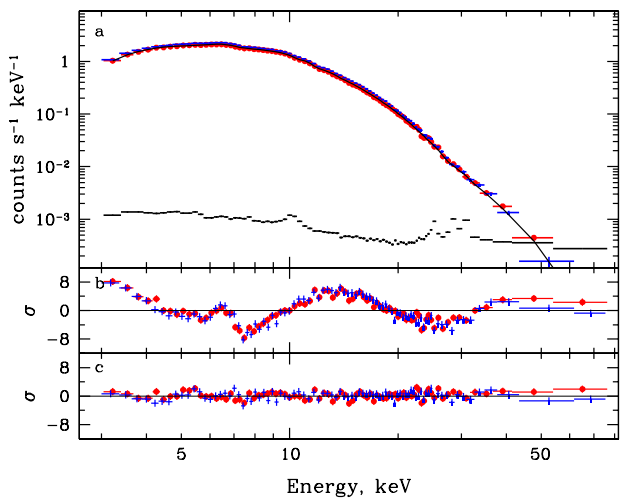}
\caption{(a) Pulse phase averaged spectrum of \2s\ obtained with the
  {\it NuSTAR} observatory. Red circles and blue crosses correspond to
  FPMA and FPMB modules data, respectively. The black line shows the
  best fit by the model consist of {\sc phabs}
\,$\times$\, ({\sc bbodyrad} \,+\, {\sc cutoffpl}\,+\, {\sc gauss})
  \,$\times$\, {\sc cyclabs}. (b) Residuals from the {\sc phabs}
\,$\times$\, {\sc
    cutoffpl} continuum model. (c) Residuals from {\sc phabs}
\,$\times$\, ({\sc bbodyrad} \,+\,
        {\sc cutoffpl}\,+\, {\sc gauss}) \,$\times$\, {\sc cyclabs}
        model.  Black crosses represent averaged background level.}\label{fig:avspec}
\end{figure}

In general, the spectrum of \2s\ has a shape which is typical for
accreting X-ray pulsars and demonstrate an exponential cutoff at high
energies (Fig.\,\ref{fig:avspec}a).  Therefore we initially
approximated it with the cutoff power-law model ({\sc phabs}
\,$\times$\, {\sc cutoffpl} in the {\sc XSPEC} package). Results of
this fitting give an unacceptable value of $\chi^2=3432.1$ for 1303
d.o.f., corresponding residuals are shown in
Fig.\,\ref{fig:avspec}(b). The source and background spectra from both
FPMA and FPMB were used for simultaneous fitting, without coadding.
To take into account the uncertainty in their calibrations a
cross-calibration constant between the modules was added to any used
spectral model.  From this figure it is clearly seen two peculiarities
in the spectrum -- an excess at low energies and depression between 20
and 30 keV. An inclusion to the model of an additional black body
component ({\sc bbodyrad} model in the {\sc XSPEC} package) with the 
temperature of about 1 keV improves significantly
the fit, but the value of $\chi^2=1726.7$ for 1300 d.o.f. is still
quite large and associated with the deficit of photons between 20 and
30 keV. Therefore at next step we introduced to the model an
absorption component in the form of the {\sc cyclabs} model in the
{\sc XSPEC} package.  It led to the following improvement of the
$\chi^2$ value to 1306.5 for 1297 d.o.f.  The corresponding residuals
are shown in Fig.\,\ref{fig:avspec}(c), the line centroid energy is
$\simeq23.5$ keV.

It is worth noticing that the appearance of this feature is not a
consequence of the combination of some particular continuum models. We
investigated several other spectral models or their combinations to
describe the continuum ({\sc powerlaw} \,$\times$\,{\sc highecut}, {\sc bbodyrad}
\,+\, {\sc bknpower}, {\sc compTT}, {\sc bbodyrad} \,+\, {\sc compTT},
{\sc compTT}\,+\, {\sc compTT}) and found that:
1) all of them approximate it insufficiently well; 2)
residuals for all of them demonstrate a prominent absorption-like
feature in the 20--30 keV energy range. Thus we can conclude that this
absorption feature is real. We interpret this feature
as a cyclotron absorption line. Usually two different models,
{\sc cyclabs} and {\sc gabs} in the {\sc XSPEC} package, are used to
describe such a feature. Both models adequately approximate an
absorption line, but the cyclotron line energy derived from the {\sc
  cyclabs} model is systematically lower (by $\sim$1--3 keV) than the
energy derived from the {\sc gabs} model \citep[see, e.g.,][]{mih95,tsy12,2015MNRAS.448.2175L}.

Additionally, the prominent emission feature near 6.4 keV is clearly detected in
the source spectrum. This feature is associated with the fluorescent
iron emission line and often observed in spectra of BeXRPs. To take it
into account we added to the model a corresponding component in the
Gaussian form.

Thus the final spectral models which were used for the approximation
of the \2s\ spectrum in the {\sc XSPEC} package are: {\sc phabs}
\,$\times$\, ({\sc bbodyrad} \,+\, {\sc cutoffpl}\,+\, {\sc gauss})
\,$\times$\, {\sc cyclabs} (Model I) and {\sc phabs} \,$\times$\,
({\sc bbodyrad} \,+\, {\sc cutoffpl}\,+\, {\sc gauss}) \,$\times$\,
  {\sc gabs} (Model II). Their best fit parameters are summarized in
  Table \ref{tablspec}. Note that the normalization of the black body
  component is
  proportional to the surface area and depends on the radius of the
  emission region $R_{\rm bb}$ and distance to the source $d$ as
  $A=(R_{\rm bb}/d_{10})^2$, where $R_{\rm bb}$ is expressed in km and
  $d_{10}$ -- in units of 10 kpc.

We are not able to detect the second harmonics of the cyclotron line 
with the existing data due to the instrumental response drop-off at high 
energies and high level of the background (see Fig.\,\ref{fig:avspec}a). 
The $3\sigma$ upper limit on the optical depth of the second harmonics 
is 1.4 assuming its width being 10 keV. More observational data or other
instruments are needed to make a final conclusion about its presence in 
the spectrum.

\begin{table}
\caption{Best-fit parameters of the \2s\ spectrum.}\label{tablspec}
\begin{tabular}{lll}
\hline
\hline
Parameter & Model I  & Model II  \\

\hline
$const$                             & $1.015\pm0.002$ & $1.015\pm0.002$ \\
$N_{\rm H}$, $10^{22}$ cm$^{-2}$       & 2.3  &  2.3  \\
$kT_{\rm BB}$, keV                  & $0.94\pm0.02 $  & $0.94\pm0.02$ \\
$A_{\rm BB}$\a                        & $18.7\pm1.3$    & $19.54\pm1.3$ \\
Photon index                        & $-0.53\pm0.06$  & $-0.66\pm0.08$ \\
$E_{\rm cut}$, keV                      & $5.59\pm0.17$   & $5.06\pm0.10$ \\
$\tau_{\rm cycl}$                       & $0.60\pm0.05$   & $8.28\pm1.26$ \\
$E_{\rm cycl}$, keV                     & $23.46\pm0.35$  & $27.34\pm0.38$  \\
$\sigma_{\rm cycl}$, keV                & $10.8\pm1.0$    & $6.44\pm0.45$    \\
$E_{\rm Fe}$, keV                       & $6.45\pm0.04$   & $6.45\pm0.04$  \\
$\sigma_{\rm Fe}$, keV                  & $0.41\pm0.06$   & $0.42\pm0.06$  \\
$EW$, eV                            & $87\pm9$        & $89\pm10$ \\
Flux (3--79 keV)\b                   & $1.025\pm0.016$ & $1.025\pm0.016$ \\
Flux (3--20 keV)\b                   & $0.880\pm0.017$ & $0.880\pm0.017$ \\
$\chi^2$ (d.o.f.)                    & $1306.5 (1297)$ & $1301.5 (1297)$  \\

\hline
\end{tabular}
\vspace{3mm}

\begin{tabular}{ll}
\a & normalization of the black body component (see text) \\
\b & in units of $10^{-9}$ \flux \\
\end{tabular}
\end{table}

\subsubsection{Pulse-phase resolved spectroscopy}

\begin{figure}
\includegraphics[width=0.98\columnwidth,bb=50 150 460 685]{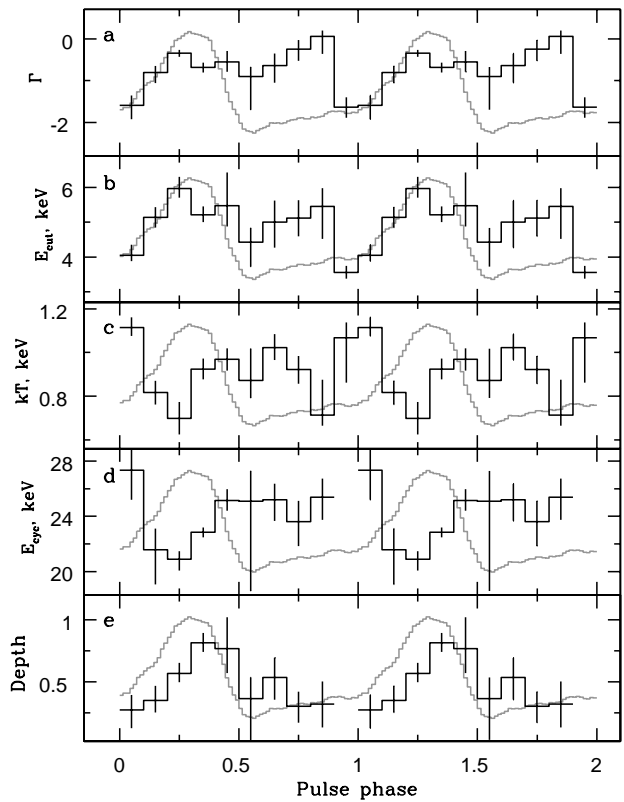}
\caption{Spectral parameters of the best-fit model (Model I, see text)
  as a function of pulse phase. The black histogram in the panels
  represents: (a) photon index, (b) cutoff energy, (c) black body
  temperature, (d) cyclotron line energy, (e) cyclotron line
  depth. Grey line in each panel shows pulse profile in wide energy
  range. The cyclotron line is not detected in the 10th bin.}\label{fig:resspec}
\end{figure}

\begin{figure}
\includegraphics[width=0.98\columnwidth,bb=45 220 555 695]{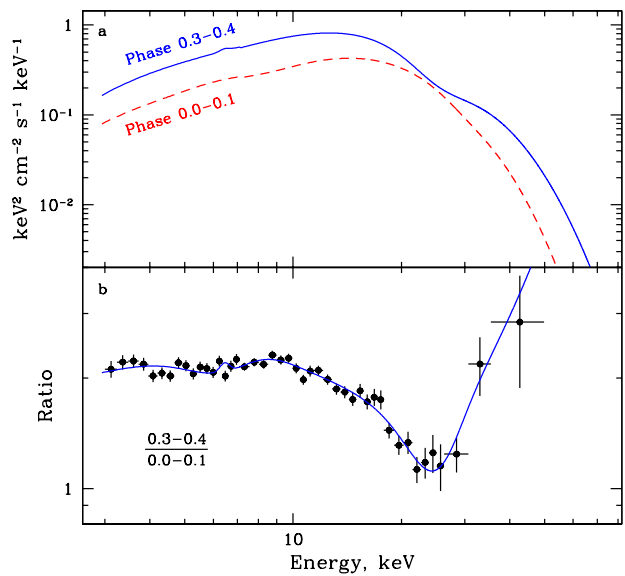}
\caption{(a) Best fit spectral models for two different phase
  bins. Red dashed line and blue solid line represent phases 0.0--0.1
  and 0.3--0.4, correspondingly. Their ratio is shown in panel
  (b). Prominent absorption-like feature is seen between 20 and 30
  keV. }\label{fig:sprat}
\end{figure}

As it was mentioned above, spectra of X-ray pulsars can vary
significantly over the pulse. The observed variations of spectral
parameters can give information about changes of the physical
conditions or parameters of the emission regions near the neutron
star. In order to study their evolution over the pulse period in the
case of \2s\ we performed a pulse-phase resolved spectroscopy. The
spin period was divided into 10 equal phase bins. The pulse profile
has a more or less smooth shape (Fig.\,\ref{fig:pprof}), therefore
such a division allows us both to obtain a good statistic for each
spectrum and to trace well evolution of spectral parameters.

To describe phase-resolved spectra we used the same spectral model as
for the analysis of the averaged spectrum (with {\sc cyclabs}
prescription for the cyclotron line; Model I). It is necessary to note
that the width of the cyclotron line cannot be firmly determined in
some phase bins due to insufficient photon statistics. Therefore, its
value was fixed at 8 keV -- the value which was measured in phase bins
near the pulse maximum with the better statistics.

The results of the pulse phase-resolved spectroscopy are shown in
Fig. \ref{fig:resspec}.  The average pulse profile of \2s\ across the
entire {\it NuSTAR} energy range is presented in each panel to
visualize better the variations of spectral parameters with the
phase. The photon index is varying quite significantly (from --2 to 0)
whereas the cutoff energy is less variable staying in the range between
4 and 6 keV. We can mention also a possible tentative correlation
between these parameters. The temperature of the black body component
is around $\simeq1$ keV and is quite stable over the pulse with some
variations from $\sim1.1$ to $\sim0.7$ keV in first three bins.

Variations of the cyclotron line energy and the line depth are the
most interesting and important for us, as both of them are changing
significantly over the pulse. In particular, the line centroid energy
has minimum ($E_{\rm cycl}\simeq21$ keV) near the pulse profile
maximum, whereas the line depth has maximum near this phase. In
general, the line energy $E_{\rm cycl}$ is varying between $\simeq21$
and $\simeq27$ keV (using {\sc cyclabs} model for the cyclotron line),
that is the main reason for the significant broadening of the line
measured in the averaged spectrum ($\sim11$ keV). Note that the
described behaviour of the cyclotron line parameters over the pulse is
very similar to the observed one in well studied BeXRP V\,0332+53 when
this object had nearly the same luminosity $L\simeq6\times10^{37}$
\lum\ \citep{2015MNRAS.448.2175L}. From another side this behaviour is
opposite to what observed in sub-critical pulsars due to different
beaming properties \citep[see, e.g., ][for the X-ray pulsar Her
  X-1]{2008A&A...482..907K}.

To demonstrate that the observed spectral variations are real ones we
show the ratio of two spectra obtained in the first (minimal line depth)
and forth (maximal line depth) phase bins in Fig. \ref{fig:sprat}. The
upper panel demonstrates corresponding models for these spectra. The
ratio of the observed spectra is shown in the bottom panel with filled
circles. The solid line in this panel represents the ratio of the
corresponding models. A strong absorption like feature, caused by
difference in line depths for these phase bins is clearly seen in this
figure between 20 and 30 keV.

The contribution of different continuum components (black body and
power-law with cutoff) to the total source luminosity demonstrates a
variability of their ratio over the pulse. Particularly, the black
body flux remains virtually constant at the level of
$\sim9\times10^{-11}$ \flux, whereas power-law flux is dominating at
all phases and is determining the overall pulse profile shape. The
contribution of the black body emission to the total flux is varying
between $\sim5$ and $\sim15$ per cent.  Using estimation of the distance to
the source $d\sim20$ kpc (see Section 4) the measured black body flux
corresponds to the emitting area with the radius of $R\sim9$ km. This
value is comparable with the neutron star size, that means that the
black body emission can emerge from the neutron star atmosphere heated
up by the intercepted emission from the accretion column
\citep{2013ApJ...777..115P}. So big radius of the illuminated NS surface
area can explain virtual constancy of the black body emission component
over the pulse. This hypothesis can be verified in future
by a set of observations of \2s\ at different luminosities and, hence,
with different illuminated areas on the neutron star.

\section{Discussion}

In this work we presented detailed spectral and temporal analysis of
the emission from the poorly studied BeXRP \2s\ using the {\it NuSTAR}
data collected during the presumably Type II outburst in 2015. The
energy spectrum of the source cannot be fitted satisfactory with any
simple continuum model but requires inclusion of the absorption
feature centered at $\sim23.5$ keV.

This absorption component has a clear physical meaning and represents
the cyclotron absorption line known to be main evidence for the strong
magnetic field on the neutron star surface
\citep{1974A&A....36..379G}. The absorption line in the \2s\ spectrum
can be fitted both by the Gaussian ({\sc gabs} model) or Lorenzian
({\sc cyclabs} model) profiles with approximately the same
quality. The line energy ($\simeq27.3$ keV for {\sc gabs} and
$\simeq23.5$ keV for {\sc cyclabs}) corresponds to the magnetic field
strength at the neutron star surface $B\sim3\times10^{12}$ G after the
correction for the gravitational redshift.

The existence of the cyclotron feature is confirmed not only by the
spectral fitting, but also by temporal properties of the source. Particularly,
the pulsed fraction dependence on the energy has a broad feature (local
decrease) around 23 keV (see Fig. \ref{fig:ppfr}) that coincides with
the position of CRSF in the source spectrum. Such non-monotonic
dependencies of the pulsed fraction on energy were observed in several
X-ray pulsars with cyclotron line in their spectra
\citep{2009AstL...35..433L,2009A&A...498..825F} and even were proposed
as a tool to search for CRSFs.

Further evidence for the cyclotron absorption feature comes from the
behaviour of the pulse profile with energy. A phase lag around
the spectral feature can be clearly seen in Figs \ref{fig:pprof} and
\ref{fig:2dpprof} expressed in a wave-like structure near 20--30
keV. Such behaviour has been shown to be typical for a few others
transient BeXRPs \citep{2006MNRAS.371...19T, 2007AstL...33..368T,2015MNRAS.454..741I} and
can be explained by a natural assumption of an energy-dependent
beaming of the radiation from the emitting region
\citep{2011A&A...532A..76F,2014A&A...564L...8S}.

The knowledge of the magnetic field strength gives us an opportunity to
estimate the absolute value of the mass accretion rate needed to provide
the spin-up rate measured by {\it Fermi}/GBM at the moment of our {\it NuSTAR}
observation $\dot P=-(7.5\pm0.9)\times10^{-10}$ s s$^{-1}$. For that
we used the accretion torque theories developed by different authors
\citep{1979ApJ...234..296G,1995ApJ...449L.153W,2007ApJ...671.1990K}. These
models explain the observed spin-up/down rate as a function of the
neutron star parameters and depend on physics of ``accretion disc --
magnetosphere'' interaction. For our calculations we used the
NS magnetic dipole moment $\mu=1.5\times10^{30}$ G cm$^3$, derived from
the measured value of the magnetic field strength, $1.4M_{\odot}$ and 10
km as the neutron star mass and radius, respectively, keeping the mass
accretion rate as a free parameter.

Type II outbursts in BeXRPs are usually accompanied by the formation
of the temporary accretion disk around the neutron star \citep[see,
  e.g.,][and references therein]{2011Ap&SS.332....1R} revealing itself
in a strong spin-up rate and properties of the noise power spectrum
\citep{2009A&A...507.1211R}. Therefore, as a first approximation we
used eq. (15) from \cite{1979ApJ...234..296G} to estimate the
bolometric luminosity $L=(7.6\pm0.9)\times10^{37}$ \lum~ needed to
support the measured spin-up rate. Given the flux from \2s\ during the
{\it NuSTAR} observation $F=1.05\times10^{-9}$ \flux~ the distance
to the system can be estimated as $d\sim22$ kpc.

The torque model by \cite{1979ApJ...234..296G} is not the only one developed for the disc
accretion. In all these models the total torque can be expressed in
the form $N_{\rm tot}=n(\omega_s)N_{\rm acc}$. The only parameter of the dimensionless angular momentum
$n(\omega_{\rm s})$ is so called fastness parameter $\omega_{\rm
s}=(r_{\rm m}/r_{\rm co})^{3/2}$, where $r_{\rm m}$ and $r_{\rm co}$
are magnetospheric and corotation radii, correspondingly. We used different
prescriptions for the dimensionless angular momentum $n(\omega_s)$
which describes physical properties of the accretion disc interaction
with the magnetosphere. Particularly, utilizing approaches from
\cite{2007ApJ...671.1990K} and \cite{1995ApJ...449L.153W} we
got distances $d\sim17$ kpc and $d\sim22$ kpc, correspondingly. Such
not large dispersion of the distances derived from different models is
due to the source being far from the spin equilibrium, where
the difference in the above-mentioned models is maximal \citep[see, e.g.,
  Fig. 2 from][]{2015arXiv150708627P}.

Radius of the magnetosphere in these models is assumed to be a
fraction of the Alfv$\acute{\rm e}$n radius $r_{\rm m}=\xi r_{\rm
  A}$. In the calculations above we assumed the parameter $\xi=0.5$,
however its exact value is not known and supposed to be between 0.5
and 0.7 \citep[see, e.g.,][and references
  therein]{2015arXiv150708627P}. To estimate a possible influence of
this parameter for our results we recalculated distances with
$\xi=0.7$. This affected the derived values very insignificantly
shifting the distance estimations based on models by
\cite{1979ApJ...234..296G} and \cite{1995ApJ...449L.153W} to $\sim$24
and $\sim$21 kpc, respectively. These estimations make \2s\ one of the
most distant high-mass X-ray binary in the Galaxy
\citep{2013MNRAS.431..327L}, putting it to the opposite side of the
Milky Way. It is important to note, that the distance estimations
based on the temporal properties of X-ray pulsars have quite large
systematic uncertainties (of the order of 15--20 per cent) due to
model dependency, an unknown efficiency of the accretion and effects
of a possible emission beaming, therefore should be considered with
the caution. Nevertheless, a large distance to the source ($>$15 kpc)
is estimated from the optical data as well \citep{lut_chasalt}.

The dispersion of the estimated distance values can be considered as a
systematic uncertainty of this method. It is interesting to note, that
if our estimations of distance ($20\pm4$ kpc) are correct then
\2s\ exceeds the so-called critical luminosity above which the
accretion column begins to grow above the neutron star surface
\citep{1976MNRAS.175..395B}. The value of the critical luminosity as
well as conditions for growing of the accretion column are still under
debates and depend on the physical models \citep[see, e.g.,][and
  references therein]{2015MNRAS.447.1847M}. A set of observations in
different intensity states during the outburst is needed to verify
this hypothesis by, e.g., observing of the anti-correlation between
the cyclotron energy and source luminosity, as seen in at least one
bright transient X-ray pulsar V\,0332+53
\citep{1998AdSpR..22..987M,2006MNRAS.371...19T,2010MNRAS.401.1628T}.
Presence of such an anti-correlation in another similar
source, 4U\,0115+63, is still under debates
\citep{2006ApJ...646.1125N,2007AstL...33..368T,2013A&A...551A...6M,2013AstL...39..375B}.

\section{Conclusion}

The recent outburst from BeXRP \2s\ was only the third
transient event when the source came into the view of X-ray
instruments. Thanks to the {\it NuSTAR} wide energy coverage and high
sensitivity we were able to discover a cyclotron absorption line with
centroid energy $E_{\rm cyc}=23.5\pm0.4$ keV, corresponding to the neutron
star magnetic field strength $B\sim3\times10^{12}$~G typical for the
known X-ray pulsars \citep{2015arXiv150503651W}. The presence of the
cyclotron line in the source spectrum is also supported by the
behaviour of the pulse profile and pulsed fraction with the energy.

The pulse-phase resolved spectroscopy revealed significant variations
of the cyclotron line parameters over the pulse. Particularly, the
line centroid energy is anti-correlating with the intensity, whereas
the line depth shows a correlation.

Thanks to the {\it Fermi}/GBM data we were able to substantially
improve the orbital parameters of the system.  The intrinsic spin
period value and its evolution observed by {\it Fermi}/GBM and {\it
  NuSTAR} during the current outburst are similar to those measured by
the {\it RXTE} observatory during the previous outburst
\citep{2012MNRAS.423.3352P}.  Virtual constancy of the period since
the source discovery in 1975 and a significant spin-up observed during
both outbursts (up to $\dot P \sim-10^{-9}$ s s$^{-1}$) implies action
of deceleration torques between outbursts. Taking into account the
last measured value of the pulse period during the previous outburst
\citep{2012MNRAS.423.3352P} and its first measured value during the
current outburst (Fig.\,\ref{fig:lc}) we can estimate roughly a
spin-down between 2007 and 2015 outbursts as $\dot P \sim
(3-4)\times10^{-11}$ s s$^{-1}$. This value is more than an order of
magnitude lower than the spin-up during the outbursts and comparable
with the spin-down observed between outbursts in other BeXRBs
\citep[see, e.g.,][]{2015MNRAS.446.1013P}.

The knowledge of the magnetic field and spin-up rate allowed us to
estimate the distance to the system $d=20\pm4$ kpc using the standard
accretion torque models. So large distance agrees well with the fact
that the optical counterpart was not directly detected so far.

\centerline{}

\section*{Acknowledgments}

This research has made
use of data obtained with {\it NuSTAR}, a project led by Caltech,
funded by NASA and managed by NASA/JPL, and has utilized the NUSTARDAS
software package, jointly developed by the ASDC (Italy) and Caltech
(USA). This research has made also by using {\it Chandra} data
provided by the Chandra X-ray Center. The publication makes use of
software provided by the Chandra X-ray Center (CXC) in the application
package CIAO. ST, AL and SM acknowledge support from Russian Science
Foundation (grant 14-12-01287). JP thanks the Academy of Finland for
financial support (grant 268740). Partial support comes from the EU
COST Action MP1304 ``NewCompStar''.

\bsp    
\label{lastpage}
\end{document}